\documentclass[%
 aip,
 pof,
 amsmath,amssymb,
 reprint, %
]{revtex4-2}

\usepackage{graphicx}
\usepackage{dcolumn}
\usepackage{bm}
\usepackage{upgreek} 
\usepackage{subcaption}
\usepackage{xcolor} 
\usepackage[normalem]{ulem} 

\usepackage[utf8]{inputenc}
\usepackage[T1]{fontenc}
\usepackage{mathptmx}
\usepackage{etoolbox}

\makeatletter
\def\@email#1#2{%
 \endgroup
 \patchcmd{\titleblock@produce}
  {\frontmatter@RRAPformat}
  {\frontmatter@RRAPformat{\produce@RRAP{*#1\href{mailto:#2}{#2}}}\frontmatter@RRAPformat}
  {}{}
}%
\makeatother
\begin{document}

\title[Power law singularity for cavity collapse in a compressible Euler fluid with Tait-Murnaghan EoS]{Power law singularity for cavity collapse in a compressible Euler fluid with Tait-Murnaghan equation of state}
\author{Daniels Krimans}
\email{danielskrimans@physics.ucla.edu}
\affiliation{Physics and Astronomy Department, University of California Los Angeles, Los Angeles, California 90095, USA}
\author{Seth Putterman}
\affiliation{Physics and Astronomy Department, University of California Los Angeles, Los Angeles, California 90095, USA}
\date{\today}

\begin{abstract}
Motivated by the high energy focusing found in rapidly collapsing bubbles that is relevant to implosion processes that concentrate energy density, such as sonoluminescence, we consider a calculation of an empty cavity collapse in a compressible Euler fluid. We review and then use the method based on similarity theory that was previously used to compute the power law exponent $n$ for the collapse of an empty cavity in water during the late stage of the collapse. We extend this calculation by considering different fluids surrounding the cavity, all of which are parametrized by the Tait-Murnaghan equation of state through parameter $\gamma$. As a result, we obtain the dependence of $n$ on $\gamma$ for a wide range of $\gamma$, and indeed see that the collapse is sensitive to the equation of state of an outside fluid. 
\newline
\newline
\textit{This article may be downloaded for personal use only. Any other use requires prior permission of the author and AIP Publishing. This article appeared in D. Krimans and S. Putterman, "Power law singularity for cavity collapse in a compressible Euler fluid with Tait–Murnaghan equation of state", Phys. Fluids }\textbf{35}\textit{, 086114 (2023) and may be found at https://doi.org/10.1063/5.0160469.}
\end{abstract}

\maketitle

\section{Introduction}
Rayleigh \cite{rayleigh_1917} calculated that the radius $R$ of an empty cavity in an incompressible ideal fluid collapsed to zero at a finite time $t_0$ as:
\begin{equation}
\label{rayleigh_collapse_power_law}
R(t) = A(t_0 - t)^{2/5},
\end{equation}
where $A = (5/2)^{2/5} (E / 2 \pi \rho)^{1/5}$, $\rho$ is the mass density of the fluid and $E = (4 \pi/3) p_0 R_m^3$ is the energy of the fluid, which is the energy to form the initial cavity of radius $R_m$ in an otherwise stationary fluid, where $p_0$ is the ambient externally applied pressure. If there are $N$ atoms inside the cavity the collapse will arrest at a finite radius $R_c$  and at this moment the energy per particle will be $E/N$. A typical experiment \cite{Lfstedt1993TowardAH} reaches $R_m = 50$  $\upmu$m with $N = 1.2 \times 10^{10}$. Due to the finite size of the atoms the collapse arrests at $R_c \approx 0.5$ $\upmu$m and at this stagnation point the average energy delivered to each particle is $E/N \approx 25$ eV. As the emission of ultraviolet photons, with an energy of $6$ eV, can be observed from the interior of the collapsed bubble \cite{PhysRevLett.69.1182}, the Rayleigh bubble dynamics is widely regarded as providing a zeroth order picture of sonoluminescence. 
\par
As the moment of the Rayleigh collapse is approached, from Eq.~\eqref{rayleigh_collapse_power_law} the velocity of the cavity wall approaches infinity as $\dot{R}(t) = (-2A/5) (t_0 - t)^{-3/5}$. Consider again the situation where gas is contained in the cavity. On the one hand it will arrest the collapse prior to reaching zero radius. On the other hand the speed of the gas at the boundary of the cavity $r = R(t)$ can become supersonic relative to the medium in the bubble which, in contrast to the external fluid, is compressible \cite{Lfstedt1993TowardAH}. If the supersonic motion occurs well before arresting of the collapse then an imploding shock wave can form. The shock can focus to the origin, $r = 0$, independent of the presence of matter. In this case $E/N \approx 1000$ eV has been theoretically predicted \cite{PhysRevLett.70.3424}. Realization of this handover in the focusing of energy density would raise prospects for the use of acoustics to achieve thermal fusion \cite{bass_2008}. Ramsey \cite{PhysRevLett.110.154301, Ramsey_PhD} has emphasized that a small compressibility of the surrounding fluid might slow down the collapse and affect the attainment of a next stage in energy focusing. In particular, one notes that in 1960 Hunter \cite{hunter_1960} calculated that for water (which is the fluid of choice for most experiments) near the moment of the collapse:
\begin{equation}
R(t) = A_n(t_0 - t)^{n},
\end{equation}
where $n = 0.5552 \approx 5/9$. Hunter used the Tait-Murnaghan \cite{cole_1948, Murnaghan1944TheCO} form of the equation of state for the fluid's pressure $p$:
\begin{equation}
p(\rho) = B\left(\left( \frac{\rho}{\rho_0} \right)^{\gamma} - 1 \right),
\end{equation}
where $B = 3000$ atm, $\gamma = 7$, $\rho_0 = 1000$ kg/$\textrm{m}^3$. The compressibility of water slows down the collapse and preliminary analysis indicates that the handover to an imploding shock wave can be suppressed \cite{doi:10.1121/2.0000869}.
\par
\begin{figure*}
\includegraphics[width=4.8in, height=2.0in]{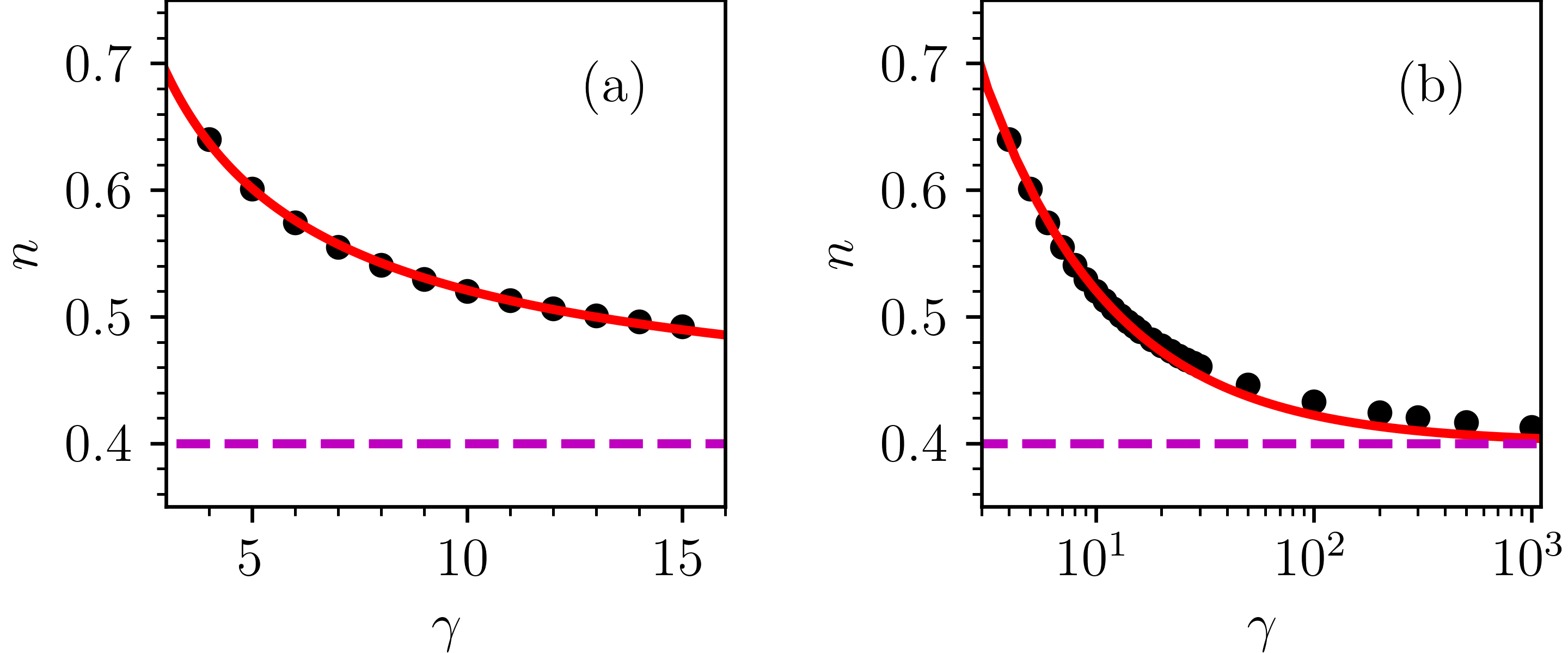} 
\caption{Predicted values of $n$ as a function of $\gamma$ (black dots), the proposed fit for the range of $\gamma$ close to the value of water that has $\gamma = 7$ of the form $n = 0.4 + a \gamma^{-b}$, where $a = 0.657424$, $b = 0.735226$ (solid red line), and the value for the incompressible fluid $n = 0.4$ (dashed magenta line). In (a), values of $\gamma$ are those close to the water that has $\gamma = 7$. In (b), a larger range of $\gamma$ is considered.}
\label{figure_results}
\end{figure*}

Even for ideal hydrodynamics the compressibility of the fluid has a strong influence on the extent to which energy density is concentrated. Motivated by this perspective we present a calculation  for $n(\gamma)$ for a wide range of materials described by different values of $\gamma$ that extends the results of Hunter for water and of Rayleigh for incompressible fluid. Key results are displayed in Fig.~\ref{figure_results}. These calculations might motivate a search for candidate liquids to achieve greater levels of energy focusing. Our calculations also provide an asymptotic limit that can be used to evaluate accuracy of more general numerical solutions, for example, simulations of all-Mach number bubble dynamics \cite{FUSTER2018752}. 

\section{Theory}
In this section, we give a summary of the methods used to perform the computations \cite{hunter_1960}. \par

We assume a spherically symmetric scenario in which a spherical empty cavity has its center placed in the origin of the coordinate system and whose radius is described by a time dependent function $R(t)$. Outside of this radius, we assume an
infinite ideal fluid which is described by mass conservation law and Euler's equation, where due to spherical symmetry fluid's only nonzero component of velocity is radial component $u$, and both radial velocity component and 
mass density $\rho$ are only functions of radial coordinate $r$ and time $t$. We do not consider an equation for entropy as we assume that the flow is homentropic. The following equations are considered for $r > R(t)$. 
\begin{equation}
\label{euler_equation}
\begin{gathered}
\rho \left( \frac{\partial u}{\partial t} + u \frac{\partial u}{\partial r} \right) = - \frac{\partial p}{\partial r},
\end{gathered}
\end{equation}
\begin{equation}
\label{mass_conservation}
\begin{gathered}
\frac{\partial \rho}{\partial t} + u \frac{\partial \rho}{\partial r} + \rho \left( \frac{2 u}{r} + \frac{\partial u}{\partial r }\right) = 0.
\end{gathered}
\end{equation}\par
To describe pressure $p$, we assume the Tait-Murnaghan equation of state, where $s$ is the specific entropy.
\begin{equation}
\label{equation_of_state}
p(\rho, s) = B(s) \left( \left(\frac{\rho}{\rho_0(s)}\right)^{\gamma} -1 \right)
\end{equation}\par
For simplicity, as the flow is homentropic, from now on we will not explicitly write dependence of $B, \rho_0$, or other entropy dependent variables on the specific entropy $s$. \par
We would like to enforce two boundary conditions for both radial velocity $u$ and pressure $p$, where one is at the interface between the empty cavity and fluid at $r = R(t)$, and the other one is far from the cavity as $r \to \infty$. For radial velocity,
we assume that the empty cavity is a free surface and that far away from the cavity the fluid is at rest. For pressure, we assume that at the surface of the cavity, pressure is zero as the cavity is empty, and far away it approaches some finite
value $p_{\infty}$. Boundary conditions are summarized next, where the dot represents the derivative with respect to the time.
\begin{equation}
\label{velocity_bc}
\begin{gathered}
u(R(t), t) = \dot{R}(t), \qquad \lim \limits_{r \to \infty} u(r,t) = 0.
\end{gathered}
\end{equation} 
\begin{equation}
\label{pressure_bc}
\begin{gathered}
p(R(t), t) = 0, \qquad \lim \limits_{r \to \infty} p(r,t) = p_{\infty}.
\end{gathered}
\end{equation} \par
Using the assumed equation of state Eq.~(\ref{equation_of_state}), it is possible to compute the speed of sound squared $c^2$ as a function of $\rho$, and change variables describing fluid from $u, \rho$ to $u, c^2$. This is convenient in order 
to apply similarity theory for the later parts of the collapse when $R(t) \to 0$, as both variables $u$ and $c^2$ can be directly compared to the velocity of cavity's wall $\dot{R}(t)$. 
\begin{equation}
\label{speed_of_sound}
\begin{gathered}
c^2 = \frac{\partial p}{\partial \rho}= \frac{B \gamma \rho^{\gamma-1}}{\rho_0^{\gamma}}, \qquad \rho= \left(\frac{\rho_0^{\gamma} c^2}{B \gamma} \right)^{1/(\gamma-1)}.
\end{gathered}
\end{equation}\par
Using the expression for $c^2$ in terms of density $\rho$ as in Eq.~\eqref{speed_of_sound}, spherical Euler's equation Eq.~(\ref{euler_equation}) and mass conservation law Eq.~(\ref{mass_conservation}) are rewritten in terms of variables $u, c^2$ as follows.
\begin{equation}
\label{euler_equation_2}
\begin{gathered}
\frac{\partial u}{\partial t} + u \frac{\partial u}{\partial r}+ \frac{1}{(\gamma-1)}\frac{\partial c^2}{\partial r} = 0, 
\end{gathered}
\end{equation}
\begin{equation}
\label{mass_conservation_2}
\begin{gathered}
\frac{\partial c^2}{\partial t} + u \frac{\partial c^2}{\partial r} + c^2(\gamma-1)\left( \frac{2u}{r} + \frac{\partial u}{\partial r} \right) = 0.
\end{gathered}
\end{equation}\par
Using Eqs.~(\ref{equation_of_state}), (\ref{speed_of_sound}), it is possible to compute boundary conditions for $c^2$ given by constants $c_0^2, c_{\infty}^2$, from the corresponding boundary conditions for pressure as in Eq.~(\ref{pressure_bc}).
\begin{equation}
\label{speed_of_sound_bc}
\begin{gathered}
c^2(R(t), t) = \frac{B \gamma}{\rho_0} = c_0^2, \qquad \lim \limits_{r \to \infty} c^2(r,t) = c^2_{\infty}.
\end{gathered}
\end{equation}\par
Instead of solving the system of partial differential equations given by Eqs.~(\ref{euler_equation_2}), (\ref{mass_conservation_2}) which would mean that we have to supply initial conditions, we consider similarity theory that is motivated by numerical results \cite{hunter_1960}. 
As we approach the last phase of the collapse where $R(t) \to 0$, we assume that the length scale of the problem is given by $R(t)$ and the scale for velocities is given by $\dot{R}(t)$. So, we seek solutions of the following form, where
we are interested in finding functions $f$ and $g$. The goal is to reduce the problem to a system of ordinary differential equations for $f$ and $g$. 
\begin{equation}
\label{similarity_assumption}
\begin{gathered}
\frac{u(r,t)}{\dot{R}(t)} = f \left( \frac{r}{R(t)} \right), \qquad \frac{c^2(r,t)}{\dot{R}^2(t) } = g \left( \frac{r}{R(t)} \right).
\end{gathered}
\end{equation} \par
Motivated by the numerical results of the full hydrodynamic equations given in Eqs.~(\ref{euler_equation_2}), (\ref{mass_conservation_2}) in the case of water\cite{hunter_1960}, we additionally assume the power law form $R(t) = A_n(t_0 - t)^n$, where $t_0$ is a time at which collapse happens and $n$ is a power law exponent that we would like to compute. Using such power law assumption and similarity approach for $u, c^2$ as in Eq.~(\ref{similarity_assumption}), we can rewrite Eqs.~(\ref{euler_equation_2}), (\ref{mass_conservation_2}) as two coupled ordinary
differential equations for functions $f$ and $g$, where we introduce variable $x = r/R(t)$. Then, differential equations have to be solved in the range $x > 1$.
\begin{equation}
\label{diff_eq_f}
\begin{gathered}
f'(x) \left( f(x)-x \right) + \left( 1 - \frac{1}{n} \right) f(x) + \frac{g'(x) }{(\gamma-1)}= 0,
\end{gathered}
\end{equation}
\begin{equation}	
\label{diff_eq_g}
\begin{gathered}
g'(x) \left( f(x)-x \right) + 2 \left( 1 - \frac{1}{n} \right) g(x)+ g(x)(\gamma-1) \left( \frac{2f(x)}{x} + f'(x) \right)= 0.
\end{gathered}
\end{equation} \par

As these equations are ordinary differential equations, we do not have to worry about what kind of initial conditions to choose for $u, c^2$, as functions $f$ and $g$ can be solved only by the boundary conditions. To compute boundary conditions from those of $u,c^2$ given in Eqs.~(\ref{velocity_bc}), (\ref{speed_of_sound_bc}), we use definitions of $f,g$ in terms of $u,c^2$ as in Eq.~(\ref{similarity_assumption}). However, for the assumed form of $c^2$ in Eq.~(\ref{similarity_assumption}), boundary conditions cannot be satisfied. Instead, because $\dot{R}(t) \to -\infty$ as $R(t) \to 0$, to get finite speeds of sounds at boundaries, we assume that $g$ is zero at the boundaries if all we are interested in is the late stage of the collapse. 
\begin{equation}
\label{f_bc}
\begin{gathered}
f(1) = 1, \qquad \lim \limits_{x \to \infty} f(x) = 0.
\end{gathered}
\end{equation}
\begin{equation}
\label{g_bc}
\begin{gathered}
g(1) = 0, \qquad \lim \limits_{x \to \infty} g(x) = 0.
\end{gathered}
\end{equation}\par
The goal now is for each value of $\gamma$ describing the equation of state of the fluid to find the value of $n$ so that differential equations given by Eqs.~(\ref{diff_eq_f}), (\ref{diff_eq_g}) are satisfied with the appropriate boundary conditions given by Eqs.~(\ref{f_bc}), (\ref{g_bc}). 

\section{Results}
It is convenient to rewrite differential equations given by Eqs.~(\ref{diff_eq_f}), (\ref{diff_eq_g}) in the following way so that each equation contains a derivative of only one of the functions. To do this, insert the expression for $g'(x)$ from Eq.~\eqref{diff_eq_g} to Eq.~(\ref{diff_eq_f}), or insert the expression for $f'(x)$ from Eq.~\eqref{diff_eq_f} to Eq.~\eqref{diff_eq_g}. 
\begin{equation}
\label{diff_eq_f_2}
\begin{gathered}
f'(x) \left( (f(x)-x)^2 - g(x) \right) = - \left( 1 - \frac{1}{n}\right) f(x)(f(x)-x) \\
+2 \left( 1 - \frac{1}{n} \right) \frac{g(x)}{(\gamma-1)} + \frac{2 f(x)g(x)}{x},
\end{gathered}
\end{equation}
\begin{equation}
\label{diff_eq_g_2}
\begin{gathered}
g'(x) \left( (f(x)-x)^2 - g(x) \right) = \left( 1 - \frac{1}{n} \right)(\gamma - 1)f(x) g(x) \\
- 2 \left( 1 - \frac{1}{n} \right) g(x)(f(x)-x) + 2 (\gamma - 1) f(x) g(x) \frac{(x - f(x))}{x}.
\end{gathered}
\end{equation}\par
Notice that the Eqs.~(\ref{diff_eq_f_2}), (\ref{diff_eq_g_2}) are singular at $x = 1$ with the boundary conditions chosen as Eqs.~(\ref{f_bc}), (\ref{g_bc}) as at $x = 1$ it is not possible to solve for $f'(1), g'(1)$ which then would be used to 
numerically approximate values of $f,g$ for some $x > 1$. So, instead of solving equations numerically from $x = 1$, we start from $x = 1 + \varepsilon$, where $0 < \varepsilon \ll 1$. To do that, we have to understand what are the new boundary conditions at such a point. To compute them, we expand both functions in $\varepsilon$ around $x =1$ as given next, where we use boundary conditions at $x = 1$ as in Eqs.~(\ref{f_bc}), (\ref{g_bc}). 
\begin{equation}
\begin{gathered}
f(1 + \varepsilon) = f(1) + f'(1) \varepsilon + O(\varepsilon^2) = 1 + f'(1) \varepsilon + O(\varepsilon^2),  \\ f'(1 + \varepsilon) = f'(1) + f''(1) \varepsilon + O(\varepsilon^2).
\end{gathered}
\end{equation}
\begin{equation}
\begin{gathered}
g(1 + \varepsilon) = g(1) + g'(1) \varepsilon + O(\varepsilon^2) =  g'(1) \varepsilon + O(\varepsilon^2), \\ g'(1 + \varepsilon) = g'(1) + g''(1) \varepsilon + O(\varepsilon^2).
\end{gathered}
\end{equation}\par
Consider differential equations given by Eqs.~(\ref{diff_eq_f_2}), (\ref{diff_eq_g_2}) up to first order in $\varepsilon$. As equations are singular at $x = 1$, no information is obtained from the zeroth order in $\varepsilon$. However, from the first order in $\varepsilon$, it is possible to compute values of $f'(1), g'(1)$ and to approximate boundary conditions as follows.
\begin{equation}
\label{f_bc_2}
\begin{gathered}
f(1 + \varepsilon) \approx 1 + \frac{1 }{\gamma}\left(3 - 2\left(1 - \frac{1}{n}\right) - 2\gamma \right) \varepsilon,
\end{gathered}
\end{equation}
\begin{equation}
\label{g_bc_2}
\begin{gathered}
g(1 + \varepsilon) \approx \left(1 - \frac{1}{n} \right)(1 - \gamma) \varepsilon.
\end{gathered}
\end{equation}\par
\begin{figure*}
\includegraphics[width=5.4in, height=1.35in]{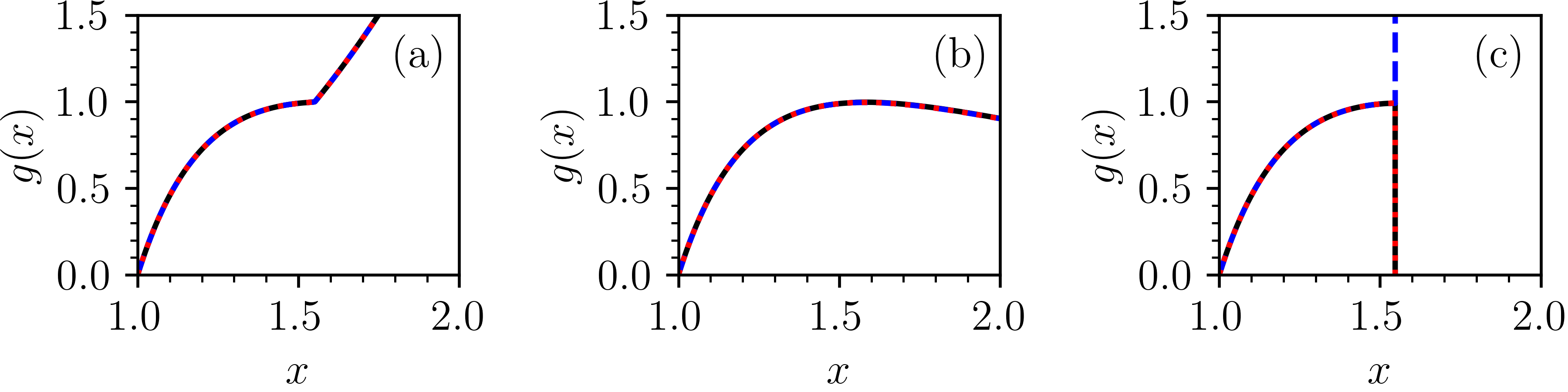} 
\caption{Numerically obtained solutions for $g(x)$ shown here from $x = 1+ \varepsilon$ until $x = 2$, where $\varepsilon = 10^{-3}$ (solid black line), $\varepsilon = 10^{-4}$ (dashed blue line), $\varepsilon = 10^{-5}$ (dotted red line). In (a), the guessed value of $n$ is smaller than the predicted one by $10^{-4}$. In (b), it is equal to the predicted value $n = 0.540799, 0.540801, 0.540801$ for $\varepsilon = 10^{-3}, 10^{-4}, 10^{-5}$, respectively. In (c), the guessed value of $n$ is larger than the predicted value by $10^{-4}$. The guessed solutions in (a) and (c) are not satisfactory because they are not smooth at around $x = 1.55$ and do not satisfy the boundary condition given in Eq.~\eqref{g_bc} as $x \to \infty$.}
\label{figure_g_plots}
\end{figure*}

For a given choice of $\gamma$, we search through values of $n$ and for each guess of $n$ we numerically integrate Eqs.~(\ref{diff_eq_f_2}), (\ref{diff_eq_g_2}) starting from $x = 1 + \varepsilon$, where boundary conditions given by Eqs.~(\ref{f_bc_2}), (\ref{g_bc_2})
are used, until $x = 5$. The maximum value of $x$ is chosen from practical considerations as then it is clear whether boundary conditions as $x \to \infty$ given by Eqs.~(\ref{f_bc}), (\ref{g_bc}) are satisfied or not. If they are, we report this value of $n$ as the predicted
value for the power law exponent of the cavity wall's collapse. Value of $\varepsilon$ used in numerical calculations is $\varepsilon = 10^{-3}$. It was checked that if this value is taken to be smaller then the results do not change significantly. For example, if $\gamma = 8$, then predicted values for $\varepsilon = 10^{-3}, 10^{-4}, 10^{-5}$ are $n = 0.540799, 0.540801, 0.540801$, respectively. As an example of a search of $n$ for $\gamma = 8$, consider results in Fig.~\ref{figure_g_plots} which show how solutions of the function $g$ look like if $n$ is smaller, equal, or greater than the predicted value. 

First, we consider obtained results for $n$ as a function of $\gamma$, where a range of $\gamma$ is chosen to be close to the value for water, $\gamma = 7$. We predict that the majority of materials have $\gamma$ values close to the one of water, so for this range, we propose the following fit that might be useful for practical applications, $n = 0.4 + a\gamma^{-b}$, where $a = 0.657424$, $b = 0.735226$. For this range of $\gamma$, the results are shown in Fig.~\ref{figure_results}(a). We have checked that our calculation for the case where $\gamma = 7$ that corresponds to the water agrees with the obtained value of $n$ of Hunter \cite{hunter_1960}. However, the described method can also be used to compute $n$ for large values of $\gamma$, results for which are given in Fig.~\ref{figure_results}(b). We see that as $\gamma \to \infty$, values indeed approach the incompressible limit $n = 0.4$ \cite{rayleigh_1917}.

\section{Conclusion}
Sonoluminescence is a phenomenon whereby a hydrodynamic singularity causes energy density to be concentrated by many orders of magnitude \cite{doi.org/10.1038/352318a0}. The irreducible and fundamental case in which this singularity is realized is when an empty cavity is surrounded by an ideal Euler fluid. Due to the presence of a singularity many additional physical factors are excited into action. These include surface tension, viscosity, thermal conduction, vapor pressure and other gases that may be present in the cavity. Here we have endeavored to study the general ideal Euler limit, which is the source of the singularity, by including the compressibility of the fluid. We find a spectrum of dynamic singularities that are determined by the thermodynamic equation of state. Our results agree with previous work of Rayleigh for the incompressible fluid \cite{rayleigh_1917} when $\gamma \to \infty$, and with Hunter \cite{hunter_1960} for the case of water where $\gamma = 7$.
From the obtained quantitative results, we see that in the late stages of collapse, the collapse is faster if $\gamma$ is larger. Therefore, the selection of materials with high $\gamma$ will facilitate the attainment of higher levels of focusing of energy density in a bubble. Using the proposed fit of our numerical data, one can readily use our results for practical applications. Our numerical results can also be used to check two fluid numerical solvers that use compressible Euler’s hydrodynamic equations by comparing the full solutions to the asymptotic empty cavity result provided in this paper. \par

Gas or vapor inside the cavity will influence the final stages of collapse. Furthermore, excitation of the gas by the collapsing cavity provides an experimental marker for this singular process. Investigation of this phenomenon is left for future papers.

\begin{acknowledgments}
This research has been funded by the DoD via HQ0034-20-1-0034 and the AFOSR under FA9550-22-1-0425 (20223965). We thank Steven Ruuth and Seth Pree for many valuable discussions.
\end{acknowledgments}



\bibliography{Power_law_singularity_for_cavity_collapse}

\end{document}